\documentclass[11pt]{article}
\usepackage{marginnote}

\usepackage{color}
\usepackage{amsmath}
\usepackage{amsthm}
\usepackage{comment}
\usepackage{marginnote}
\allowdisplaybreaks

\usepackage{array}
\usepackage{tabularx}
\usepackage{appendix}
\usepackage{makecell}
\usepackage{graphicx}
\usepackage{amssymb}
\usepackage[T1]{fontenc}
\usepackage[utf8]{inputenc}
\usepackage[raggedright]{titlesec}
\usepackage{blindtext}
\usepackage{commath}
\usepackage{cite}
\usepackage{caption}
\usepackage[textwidth = 7in]{geometry}
\usepackage{mathtools}
\usepackage[dvipsnames]{xcolor}
\usepackage{enumitem}
\usepackage{amsmath}
\usepackage{geometry}
\usepackage{lipsum}  
\usepackage{thmtools}
\usepackage{setspace}
\titleformat{\paragraph}[hang]{\normalfont\normalsize\bfseries}{\theparagraph}{1em}{}
\titlespacing*{\paragraph}{0pt}{3.25ex plus 1ex minus .2ex}{0.5em}
\makeatletter
\DeclareRobustCommand{\change}{%
	\@bsphack
	\leavevmode
	\color{magenta}%
	\@esphack
}
\DeclareRobustCommand{\stopchange}{%
	\@bsphack
	\normalcolor
	\@esphack
}
\makeatother

\stepcounter{secnumdepth}
\stepcounter{tocdepth}

\usepackage{hyperref}
\hypersetup{
  colorlinks   = true, 	
  urlcolor     = blue, 	
  linkcolor    = blue, 	
  citecolor   = magenta 	
}

\title{
	\vskip-1.3cm
Exact Floquet solutions in a Parity-Time-Symmetric Rabi Model\\
}

\author{
	M. Baradaran$^1$\footnote{marzieh.baradaran@uhk.cz, ORCID: \href{http://orcid.org/0000-0002-8455-9973}{0000-0002-8455-9973}}, 
	D. Braak$^2$\footnote{daniel.braak@uni-a.de, ORCID: \href{https://orcid.org/0000-0002-7294-4401}{0000-0002-7294-4401}}, 
	L.M. Nieto$^3$\footnote{luismiguel.nieto.calzada@uva.es, ORCID: \href{http://orcid.org/0000-0002-2849-2647}{0000-0002-2849-2647}}, 
	and 
	S. Zarrinkamar$^{3,4}$\footnote {saber.zarrinkamar@uva.es, ORCID: \href{http://orcid.org/0000-0001-9128-4624}{0000-0001-9128-4624}}
	\\  [1ex]
	\small
	$^1$\,Department of Physics, Faculty of Science, University of Hradec Kr\'alov\'e,\\
	\small
	Rokitansk\'eho 62, 500 03 Hradec Kr\'alov\'e, Czechia
	\\ [1ex]
	\small
	$^2$\,Institute of Physics, University of Augsburg, 86135 Augsburg, Germany
	\\ [1ex]
	\small
	$^3$\,Departamento de F\'{\i}sica Te\'{o}rica, At\'{o}mica y \'{O}ptica, and Laboratory for Disruptive \\ 
		\small
		Interdisciplinary Science (LaDIS), Universidad de Valladolid, 47011 Valladolid, Spain
	\\ [1ex]
	\small
	$^4$\,Departament of Basic Sciences, Garmsar Branch,
	Islamic Azad University, Garmsar, Iran
}

\begin{document}
	
	\maketitle

\begin{abstract}

It is shown that a semiclassical Rabi model with parity-time (PT) symmetry has a hidden $sl(2)$ symmetry and hence possesses quasi-exact solutions.
These are located precisely at the exceptional points of the spectrum, the boundaries of the PT-symmetric phase. The corresponding constraints on the model parameters can be interpreted as a resonance relationship between the constant and periodic driving terms. 

\end{abstract}


	\section{Introduction}
The concept of parity-time symmetry \cite {Bender1998, Bender2024} and its experimental verification in optics \cite {Ruter, Konotop} has motivated a wide range of research on this topic. In recent years, experimental evidence for parity-time symmetries has been found in several fields, including microresonators \cite {NatureChang}, microcavities \cite {NaturePeng}, quantum critical phenomena \cite {NatureAshida}, quantum circuits \cite {NatureAssawaworrarit}, quantum walks \cite {NatureXiao}, photonic crystals \cite {NatureWeimann}, sensor telemetry \cite {NatChen}, wireless power transfer \cite {NatureFan}, integrated electronics \cite {NatureCao} and photonics \cite {NatureLiu}, and photonic topological insulators \cite {NatureFritzsche}.

On the other hand, the Rabi model \cite {Rabi}, which successfully explains the simplest manifestation of light-matter interaction by coupling a two-level system to a single radiation mode, has been the focus of recent research due to its applications in various physical and interdisciplinary fields, including quantum optics, communication technology, quantum information technology, laser physics, and solid-state physics \cite{Xie 2013}.	Numerous current investigations combine the Rabi model with PT symmetry both from theoretical and experimental points of view.

	A non-Hermitian semiclassical Rabi model with an imaginary periodic driving term was studied in the Floquet picture and shown to exhibit (broken) PT symmetry \cite{Xie 2013}.
	The authors applied a perturbative approach beyond the rotating-wave approximation and showed that the non-Hermitian analogue of the Bloch-Siegert shift appears in the case of maximum PT breaking. 
	Experimental implementations include spatially modulated loss waveguides, see for example \cite {Lee}.
The stabilization of a non-Hermitian Rabi Hamiltonian via a periodic driving field and the correspondence to the band structure of a particular lattice Hamiltonian via the Floquet operator is addressed in \cite {Gong}.

	In a very recent experimental survey, PT  symmetry breaking transitions are reported using single ultracold atoms \cite {Li}.
	Anisotropic and asymmetric generalizations of the quantum Rabi model augmented with periodic driving have been analyzed in the Floquet picture \cite{Liu}. 
	A generalized adiabatic approximation has been applied to  the PT-symmetric quantum Rabi model,  and the so-called exceptional points are discussed in depth in relation to their Hermitian analogue, the Juddian level crossings \cite{Lu}. On the other hand,  multi-photon resonances, where simultaneous production of several photons can occur via the down conversion of a high frequency pump, are of great importance in the study of phase-space crystals \cite {rich periodic structure phase-space crystal}, quantum error correction \cite {quantum error correction} and superconducting cavities where notable features, including  higher order squeezing  \cite {higher order squeezing},  are observed.
	
In this work we revisit the semiclassical PT-symmetric Rabi model investigated in \cite{Xie PRA 2018} and explore the algebraic structure of the problem.
The paper is organized as follows. We first revisit the model studied in \cite{Xie PRA 2018}. We then show that, using suitable transformations, the resulting ordinary differential equation can be written in terms of $sl(2)$ generators  in an finite-dimensional representation if certain constraints on the parameters are satisfied \cite{Turbiner 88}. 
	The latter are then interpreted as resonance conditions. 
Concluding remarks are given in the last section.

	\section{ Formulation and General Solution via Lie-algebraic Approach}
	We consider the following non-Hermitian time-dependent Hamiltonian \cite {Xie PRA 2018} 
	\begin{equation}
		H=i\frac{\gamma}{2}\sigma_z+\frac{\nu(t)}{2}\sigma_x\,,
	\end{equation}
	where $\sigma_{x,z}$ are Pauli matrices acting on a pseudo-spinor $(a_1(t),a_2(t))^T$ and 
$ 
\nu(t)=\nu_0+\nu_1\cos(\omega t),
$ 
being  $\nu_1$ and $\omega$ the modulation amplitude and frequency, respectively, while $\nu_0$ represents the qubit splitting energy. In addition, $\gamma$ denotes the strength of the loss/gain term, characteristic of PT-symmetric dissipative systems.   Hence, the two-level probability amplitudes $a_1(t)$ and $a_2(t)$ satisfy the coupled first-order differential equations
	\begin{equation}
		\begin{gathered}
			i\frac{da_1(\tau)}{d\tau}=\frac{\nu(\tau)}{2\omega}a_2(\tau)+i\frac{\gamma}{2\omega}a_1(\tau),\\ 
			i\frac{da_2(\tau)}{d\tau}=\frac{\nu(\tau)}{2\omega}a_1(\tau)-i\frac{\gamma}{2\omega}a_2(\tau),
		\end{gathered}
	\end{equation}
	where we have set $\tau=\omega t$ and $\hbar=1$.
	Following \cite{Xie PRA 2018}, we introduce the combinations 
		\begin{equation}
c_1(t)=(a_1(t)+a_2(t))/2, \qquad c_2(t)=(a_1(t)-a_2(t))/2,
	\end{equation}
and  define   
		\begin{equation}
	\begin{aligned}
	&	z=e^{i\tau},\\
		& c_j(z)=e^{\frac{\nu_1}{2\omega}\cos(\tau)} z^{-\frac{\nu_0}{2\omega}} \phi_j(z),\qquad j=1,2.
	\end{aligned}
\end{equation}
	After eliminating $\phi_2(z)$, we obtain a second-order differential equation for $\phi_1(z)\equiv\phi(z)$ \cite{Xie PRA 2018}, 
	\begin{equation}\label{Rab}
		z^2\frac{d^2\phi(z)}{dz^2}+\left(\frac{\nu_1}{2\omega}z^2+(1-\frac{\nu_0}{\omega})z-\frac{\nu_1}{2\omega}\right)\frac{d\phi(z)}{dz}+\left(\frac{\gamma^2-\nu_1^2}{4\omega^2}+\frac{\nu_1}{2\omega}(1-\frac{\nu_0}{\omega})z\right)\phi(z)=0.
	\end{equation}
	This equation has the form of an eigenvalue equation $\tilde{H}\phi(z)=\lambda\phi(z)$ with fixed eigenvalue $\lambda=0$.

	We will now show that the differential operator $\tilde{H}$ in \eqref{Rab} has a ``hidden'' algebraic structure if certain conditions are met, which will be specified below.
	This means that $\tilde{H}$ can be expressed as an element of the universal enveloping algebra of $sl(2)$, acting on a \emph{finite-dimensional} representation. The general form of such a quasi-exactly solvable operator $H_{qes}$ reads \cite {Turbiner 88, Turbiner, artemio},
	\begin{equation}\label{hques1}
		H_{qes}=C_{++}{J}_n^+{J}_n^++C_{+0}{J}_n^+{J}_n^0+C_{+-}{J}_n^+{J}_n^-+  C_{0-}{J}_n^0{J}_n^-+C_{--}{J}_n^-{J}_n^-+C_+{J}_n^++C_0{J}_n^0+C_-{J}_n^-+C,
	\end{equation}
	in which the generators of $sl(2)$ are represented as first-order differential operators,
	\begin{equation}\label{GenerJJJ}
			J_n^+ =z^2\,\frac{d}{dz}-n\,z ,\qquad
			J_n^0 = z\,\frac{d}{dz}-\frac n2,\qquad
			J_n^- = \frac{d}{dz}.
	\end{equation}
	They satisfy the commutation relations
	\begin{equation}
		[J^+_n,J^-_n]=-2J^0_n,  \qquad   [J^{\pm}_n,J^0_n]=\mp J^{\pm}_n.
	\end{equation}
	For a nonnegative  integer $n$, any product of operators \eqref{GenerJJJ} leaves invariant the $(n+1)$-dimensional space of polynomials $\mathcal{P}_{n}[z]=\langle 1,z,z^2,\dots,z^n\rangle$ with maximum order $n$  because the ascending ladder operator $J_n^+$ annihilates $z^n$, which is therefore the highest weight vector of the $(2j+1)$-dimensional representation of $sl(2)$ with $j=(n-1)/2$. 
	The constant function is annihilated by $J_n^-$, which is the lowest weight vector ($n=0$ corresponds to the trivial representation with $J_0^{\pm,0}\langle1\rangle=0$). In other words, $H_{qes}$ preserves the finite-dimensional space of polynomials of the form 
	\begin{equation}\label{phi_n(z)}
		\phi_n(z)=\sum_{m=0} ^n{c_m z^m}.
	\end{equation}
	Substituting \eqref{GenerJJJ} into \eqref{hques1}, the operator $H_{qes}$ is given by a second-order differential operator with polynomial coefficients,
	\begin{equation}
		H_{qes}=P_4(z)\frac{d^2}{dz^2}+P_3(z)\frac{d}{dz}+P_2(z),
	\end{equation}
	in which
	\begin{equation}\label{generalcoefficients}
		\begin{aligned}
			&	 P_4(z)=C_{++}z^4+C_{+0}z^3+C_{+-}z^2+C_{0-}z+C_{--},\\
			&	P_3(z)=C_{++}(2-2n)z^3+\left (C_++C_{+0}\left (1-\frac{3n}{2}\right)\right)z^2+\left (C_0-nC_{+-}\right)z+\left (C_--\frac{n}{2}C_{0-} \right), \\
			&	 P_2(z)=C_{++}n(n-1)z^2+\left(\frac{n^2}{2}C_{+0}-nC_+\right)z+\left (C-\frac{n}{2}C_0 \right).
		\end{aligned}
	\end{equation}
	Now, comparing \eqref{Rab} with \eqref{generalcoefficients}, we find that if the constraint 
	\begin{equation}\label{npho}
		\nu_0=(n+1)\omega
	\end{equation} 
	holds, the equation \eqref{Rab} can be written as $\tilde H \phi_n(z)=0$ and $\tilde{H}$ given in terms of $J_n^{\pm}$:
	\begin{equation} \label {kept term}
		\tilde H={J}_n^+{J}_n^-+\frac{\nu_1}{2\omega}\left({J}_n^+-  {J}_n^-\right) 
			+  \frac{\gamma^2-\nu_1^2}{4\omega^2}.
		\end{equation}
		The condition \eqref{npho} fixing  $\nu_0$ in terms of $\omega$ and $n$ is necessary to cancel terms proportional to $z$ in \eqref{Rab} that cannot be absorbed in an expression containing $J_n^+$.
		Since $\tilde{H}$ leaves the space $\mathcal{P}_n[z]$ invariant, it can be expressed as a matrix $\tilde{H}_n$ of dimension $(n+1)\times(n+1)$. Using as basis $v_0=1,v_1=z,\ldots v_{n}=z^n$, we obtain the matrix elements of $J_n^\pm$ and $J_n^+J_n^-$ from \eqref{GenerJJJ} as
		\begin{equation}
			J^+_n(i,j)=-(n-j)\delta_{i,j+1}, \quad J^-_n(i,j)=j\delta_{i,j-1},
			\quad J^+_nJ^-_n(i,j)=(j-n-1)j\delta_{i,j},
		\end{equation}
		for $i,j=0\ldots n$.
		Therefore, $\tilde{H}_n$ is a tridiagonal matrix with elements
		\begin{equation}
			\tilde{H}_n (i,j)=\left((j-n-1)j+ \frac{\gamma^2-\nu_1^2}{4\omega^2}\right)\delta_{i,j}
			-\frac{\nu_1}{2\omega}\big((n-j)\delta_{i,j+1} +j\delta_{i,j-1}\big).
			\label{matrix}
		\end{equation}
		The condition for the existence of a non-trivial solution of $\tilde{H}_n\phi_n=0$ is $\det(\tilde{H}_n)=0$, which gives a second condition that the parameters $\nu_1,\gamma$ and $\omega$ must satisfy.
We then report explicit solutions for $n=0,1,2$ and give the constraints on the determinants for $n$ up to 5.

			\begin{itemize}
			\item  For $n=0$, that is $\frac{\nu_0}{\omega}=1$, we have $\phi_0 (z)=c_0$ for which the equation $\tilde H_0\,\phi_0(z)=0$ becomes $ (\nu_1^2-\gamma^2)c_0=0$.
			Thus, a non-trivial solution requires 
			$ 	\nu_1^2=\gamma^2,$
			meaning that the driving amplitude $\nu_1$  equals the gain/loss strength $\gamma$ (both quantities can be chosen positive without loss of generality). The condition does not depend on the driving frequency $\omega$, so this special solution does not involve a resonance phenomenon in the proper sense.
			\item  For $n=1$ ($\frac{\nu_0}{\omega}=2$), we have for $\phi_1(z)$ in the basis given above $\phi_1=(c_0,c_1)^T$ for which the equation $\tilde{H}_1\phi_1(z)=0$ reads
			\begin{equation*}\label{two photon}
				\left(\begin{array}{cc} \frac{\gamma^2-\nu_1^2}{4\omega^2} & -\frac{\nu_1}{2\omega}\\ -\frac{\nu_1}{2\omega}& \frac{\gamma^2-\nu_1^2}{4\omega^2}-1 \end{array}\right)\left(\begin{array}{c}c_0\\c_1\end{array}\right) =0,
			\end{equation*} 
			The condition $\det(\tilde{H}_1)=0$ yields,
			\begin{equation}\label{n1eq}
				(\nu_1^2-\gamma^2)^2=4\omega^2\gamma^2,
			\end{equation} 
			and the coefficients $c_0,c_1$ satisfy
			$$  \frac{c_1}{c_0 }=\frac{\gamma^2-\nu_1^2}{2\omega\nu_1}. $$
			We observe that in this case, the condition on $\nu_1$ and $\gamma$ involves the driving frequency $\omega$, so it corresponds to a resonance between the gain/loss process and the driving. From \eqref{n1eq}, explicit solutions of $\frac{\nu_1^2}{\omega^2}$ are obtained as
			\begin{equation*}
				\frac{\nu_1^2}{\omega^2}=\frac{\gamma ^2}{\omega ^2}\pm \frac{2 \gamma }{\omega }.
			\end{equation*} 

			\item   For $n=2$, meaning $\frac{\nu_0}{\omega}=3$, the matrix equation $\tilde{H}_2\,\phi_2(z)=0$ reads,
			\begin{equation*}\label{three-photon-1}
				\left(\begin{array}{ccc} 
					\frac{\gamma^2-\nu_1^2}{4\omega^2} & -\frac{\nu_1}{2\omega}&0\\
					-\frac{\nu_1}{\omega}&\frac{\gamma^2-\nu_1^2}{4\omega^2}-2&-\frac{\nu_1}{\omega}\\
					0&-\frac{\nu_1}{2\omega}&\frac{\gamma^2-\nu_1^2}{4\omega^2}-2 \end{array}\right)
				\left(\begin{array}{c}c_0\\c_1\\c_2\end{array}\right) =0.
			\end{equation*} 
			The determinant condition reads now
			\begin{equation}\label{n2eq}
				\left(\frac{\nu_1^2}{\omega^2}-\frac{\gamma^2}{\omega^2}\right)^3-16\left(\frac{\nu_1^2}{\omega^2}-\frac{\gamma^2}{\omega^2}\right)\frac{\gamma^2}{\omega^2}-64\frac{\gamma^2}{\omega^2}=0,
			\end{equation}
			and, consequently, a nontrivial solution for the matrix equation is obtained as
			\begin{equation} 
			c_1=\frac{4 \omega\nu_1   \left(\gamma ^2-\nu_1^2-8 \omega ^2\right)}{8 \omega ^2 \left(\nu_1^2-2 \gamma ^2\right)+\left(\gamma ^2-\nu_1^2\right)^2+64 \omega ^4}\,c_0,
				\quad 
				c_2=\frac{8 \omega ^2\nu_1^2 }{8 \omega ^2 \left(\nu_1^2-2 \gamma ^2\right)+\left(\gamma ^2-\nu_1^2\right)^2+64 \omega ^4}\,c_0.
			\end{equation}
			From \eqref{n2eq}, we obtain three solutions for $\frac{\nu_1^2}{\omega^2}$ in the form of Cardano’s formula, which are of the form
			\begin{align*}
				\frac{\nu_1^2}{\omega^2}&= \left(1+\frac{4 }{\xi }\sqrt[3]{\frac{2}{3}} \right) \frac{\gamma ^2}{\omega ^2}+2 \left(\frac{2}{3}\right)^{2/3} \xi ,\\[5pt]
				\frac{\nu_1^2}{\omega^2}&= \left( 1-\frac{4 }{\xi }\sqrt[3]{-\frac{2}{3}}\right) \frac{\gamma ^2}{\omega ^2}  +2 \left(-\frac{2}{3}\right)^{2/3} \xi ,\\[5pt]
				\frac{\nu_1^2}{\omega^2}&=  \left( 1+\frac{ 4 }{\xi }(-1)^{2/3}\sqrt[3]{\frac{2}{3}}\right) \frac{\gamma ^2}{\omega ^2}-2 \sqrt[3]{-1} \left(\frac{2}{3}\right)^{2/3} \xi  ,
			\end{align*} 
			in which $\xi=\sqrt[3]{\frac{9 \gamma ^2}{\omega ^2}+\frac{\sqrt{81 \gamma ^4 \omega ^2-12 \gamma ^6}}{\omega ^3}}$.
	\end{itemize}	
		
		In a similar way, one can compute the determinant conditions  for larger values of $n$. Let us summarize our results determined explicitly for $n=0,1,...,5$, or equivalently, for $\frac{\nu_0}{\omega}=1,2,...,6$\,:
\begin{eqnarray}      
				\frac{\nu_0}{\omega}=1:  && \left(\frac{\nu_1^2}{\omega^2}-\frac{\gamma^2}{\omega^2}\right)=0 ,	
				\nonumber  \\[4pt]
				\frac{\nu_0}{\omega}=2:   &&\left(\frac{\nu_1^2}{\omega^2}-\frac{\gamma^2}{\omega^2}\right)^2-4\frac{\gamma^2}{\omega^2}=0,	\nonumber  \\[4pt] 
				\frac{\nu_0}{\omega}=3: \ &&\left(\frac{\nu_1^2}{\omega^2}-\frac{\gamma^2}{\omega^2}\right)^3-16\left(\frac{\nu_1^2}{\omega^2}-\frac{\gamma^2}{\omega^2}\right)\frac{\gamma^2}{\omega^2}-64\frac{\gamma^2}{\omega^2}=0, \nonumber	\\[4pt]
				\frac{\nu_0}{\omega}=4:   &&\left(\frac{\nu_1^2}{\omega^2}-\frac{\gamma^2}{\omega^2}\right)^4-40\left(\frac{\nu_1^2}{\omega^2}-\frac{\gamma^2}{\omega^2}\right)^2\frac{\gamma^2}{\omega^2}+48\left(11\frac{\gamma^2}{\omega^2}-8\frac{\nu_1^2}{\omega^2}\right)\frac{\gamma^2}{\omega^2}-2304\frac{\gamma^2}{\omega^2}=0,  \nonumber	\\[10pt]
				\frac{\nu_0}{\omega}=5:   &&\left(\frac{\gamma ^2}{\omega ^2}-\frac{\nu_1^2}{\omega ^2}\right)^5-80 \left(\frac{\gamma ^2}{\omega ^2}-\frac{\nu_1^2}{\omega ^2}\right)^3\frac{\gamma^2}{\omega^2}+64 \left(\frac{37 \gamma ^4}{\omega ^4}-\frac{58 \gamma ^2 \nu_1^2}{\omega ^2 \omega ^2}+\frac{21 \nu_1^4}{\omega ^4}\right)\frac{\gamma^2}{\omega^2}   \label{sixCond}\\
				&& \qquad -6144 \left(\frac{5 \gamma ^2}{\omega ^2}-\frac{3 \nu_1^2}{\omega ^2}\right)\frac{\gamma^2}{\omega^2}+147456\frac{\gamma^2}{\omega^2}=0,  \nonumber
				\\[10pt]
				\frac{\nu_0}{\omega}=6:  &&  \left(\frac{\gamma ^2}{\omega ^2}-\frac{\nu_1^2}{\omega ^2}\right)^6-140 \left(\frac{\gamma ^2}{\omega ^2}-\frac{\nu_1^2}{\omega ^2}\right)^4\frac{\gamma ^2}{\omega ^2}+112 \left(\frac{69 \gamma ^2}{\omega ^2}-\frac{32 \nu_1^2}{\omega ^2}\right) \left(\frac{\gamma ^2}{\omega ^2}-\frac{\nu_1^2}{\omega ^2}\right)^2\frac{\gamma ^2}{\omega ^2}   \nonumber   \\
				&&\qquad -64 \left(\frac{3281 \gamma ^4}{\omega ^4}-\frac{4352 \left(\gamma ^2 \nu_1^2\right)}{\omega ^4}+\frac{1296 \nu_1^4}{\omega ^4}\right) \frac{\gamma ^2}{\omega ^2}+20480 \left(\frac{137 \gamma ^2}{\omega ^2}-\frac{72 \nu_1^2}{\omega ^2}\right)\frac{\gamma ^2}{\omega ^2}   \nonumber  \\
				&&\qquad -14745600\frac{\gamma ^2}{\omega ^2}  =0.  \nonumber
		\end{eqnarray}

\begin{figure}[htb]
			\centering
			\includegraphics[width=0.85\linewidth]{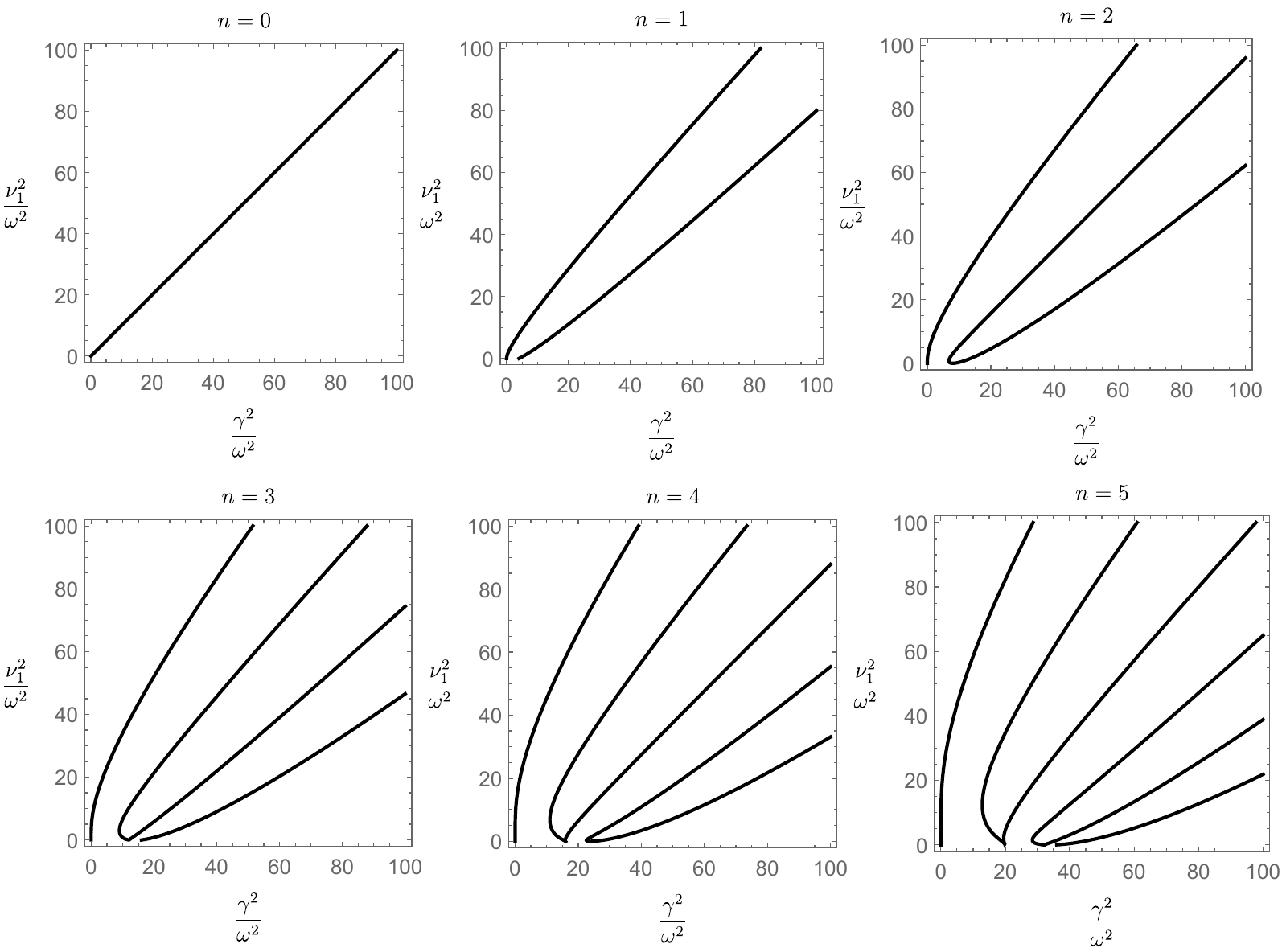}
			\caption{The $n$-photon constraints for $n=0,1,2,3,4$ and $5$.}
			\label{n012345}
		\end{figure}

\begin{figure}[tb]
				\centering
				\includegraphics[width=0.40\linewidth]{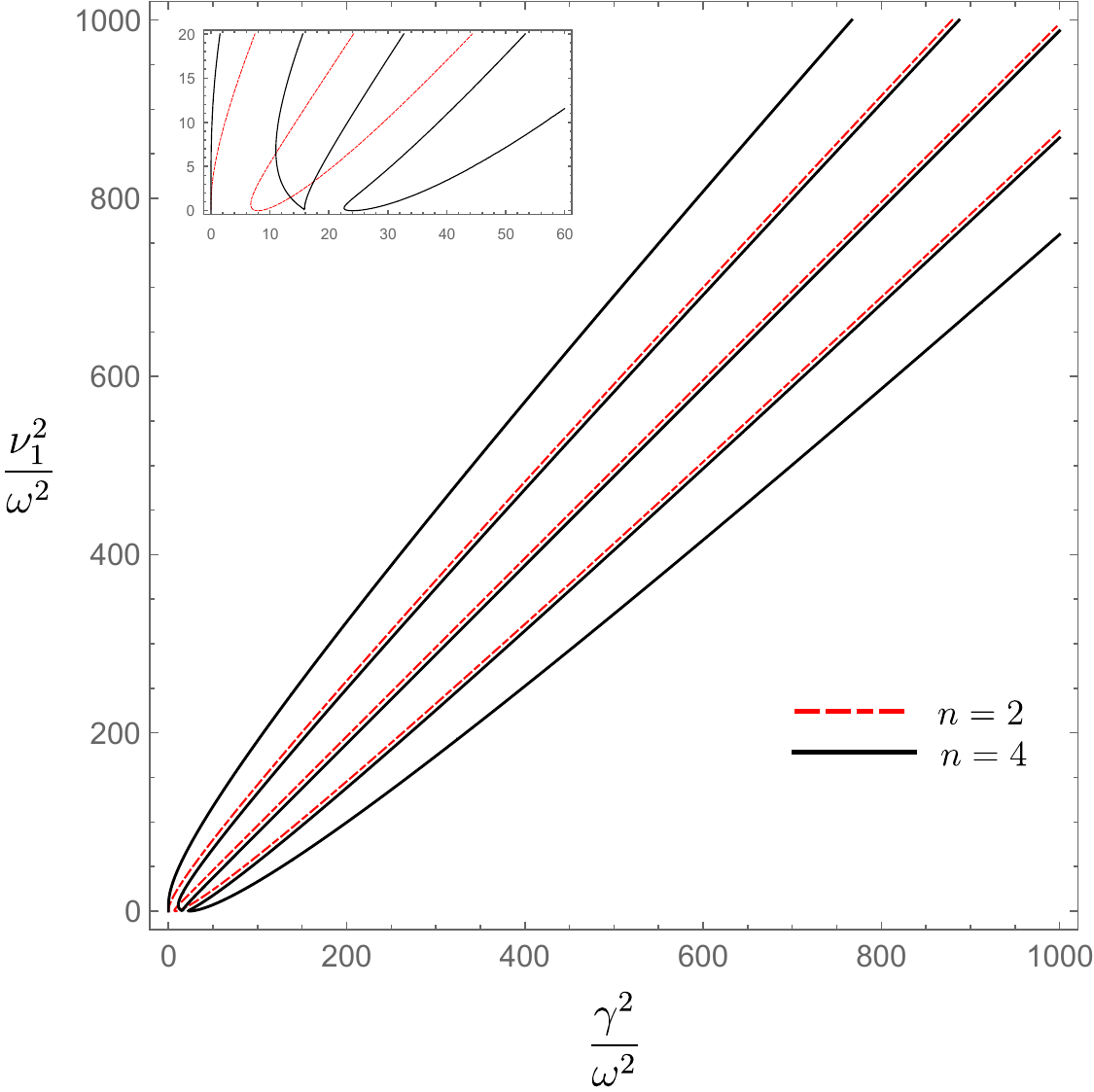} 
				\qquad
				\includegraphics[width=0.40\linewidth]{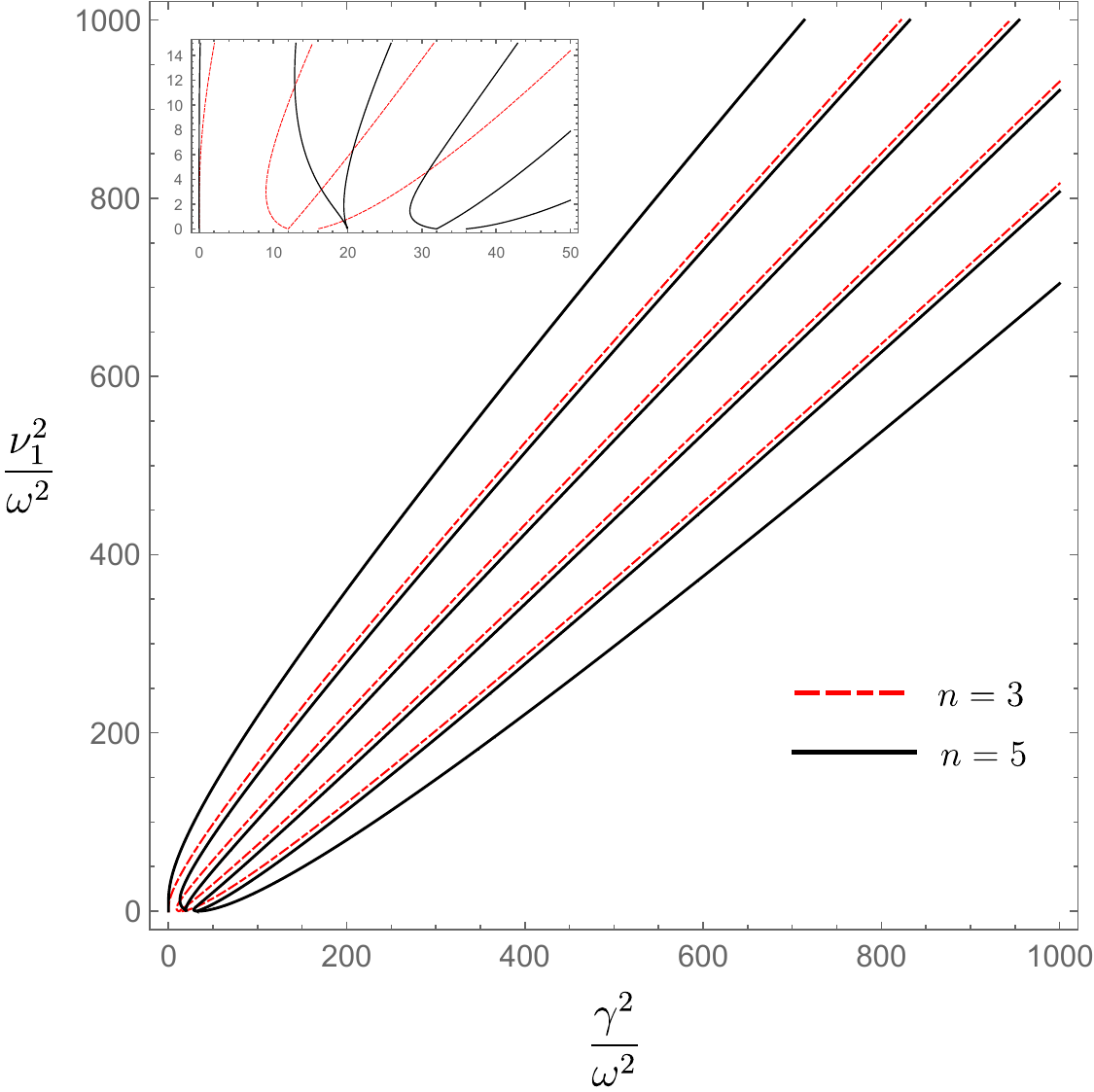} 
			   \caption{The crossing points for doublets $(n_1,n_2)=(2,4)$ on the left and $(n_1,n_2)=(3,5)$ on the right.}
				\label{n2n4n3n5}
				\end{figure}

The behavior of $\frac{\nu_1^2}{\omega ^2}$ in terms of $\frac{\gamma ^2}{\omega ^2}$, for the six conditions given in \eqref{sixCond}, is illustrated separately in Figure~\ref{n012345}. 
                We see that the general pattern of solution points starting with $n\ge1$ is similar for all $n$ with given parity (even or odd $n$) and large $\nu_1^2,\gamma^2$. This is illustrated in  Figure~\ref{n2n4n3n5} where the solution sets are compared for $(n_1,n_2)=(2,4)$ and $(3,5)$ respectively. The solution set for $n_1<n_2$ is very close to a subset of the solutions for $n_2$ if $\nu_1^2$ and $\gamma^2$ are sufficiently large. On the other hand, for small $\nu_1^2$, there are large deviations between the solution sets for $n_1$ and $n_2$ as seen in the insets of Figure~\ref{n2n4n3n5}. For $\nu_1=0$ all solutions for $\gamma^2$ are obtained by the equations
\begin{equation}\label{doublydegeneratesolutions}
  (n+1-j)j=\frac{\gamma^2}{4\omega^2}, \quad j=0,\ldots, n.
\end{equation}
This leads to doubly degenerate solutions for $j$ and $j'=(n+1-j)$ which means that for this value of the gain/loss strength $\gamma$, there are two possible Floquet solutions with $\phi_1(t)=e^{ij\omega t}$ and $\phi_1(t)=e^{i(n+1-j)\omega t}$, respectively.

\begin{figure}[htb]
	\centering
	\includegraphics[width=0.5 \linewidth]{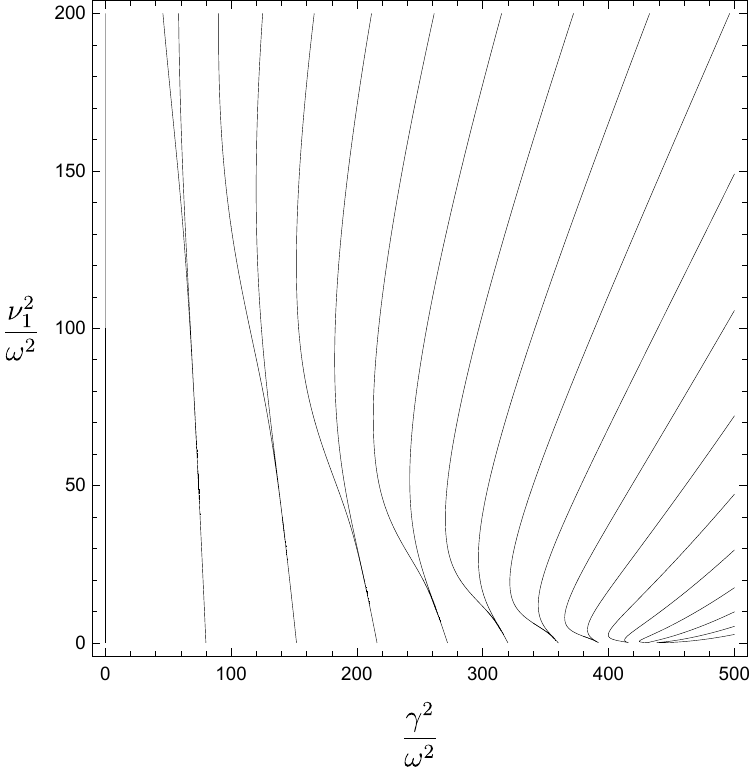}
	\caption{The $n$-photon constraint for $n=20$\,. Doubly degenerate solutions happen for $\nu_1=0$ and $\frac{\gamma ^2}{\omega ^2}=\{80,152,216,272,320,360,392,416,432,440\}$ corresponding to the pairs $j=1,\dots,n$ and $j'=(n+1-j)$ in \eqref{doublydegeneratesolutions}; the ``almost degenerate'' point is clearly visible (here) at $(\nu_1,\frac{\gamma ^2}{\omega ^2})=(0,80)$ corresponding to the pair $(j,j')=(1,20)$. }
	\label{n20degen500}
\end{figure}

                        Our numerical results suggest that these degeneracies at the trivial point $\nu_1=0$ develop into two types for small but non-zero values of $\nu_1$. In one case, the degenerate points separate quickly from each other for $\nu_1^2>0$ as seen e.g. for $n=2$ in Figure~\ref{n012345}. In the other case, the two solutions stay very close for  $\nu_1^2>0$ which entails a near degeneracy also for non-zero $\nu_1$. The ``almost degenerate'' points are most clearly pronounced for the pairs $(j,j')=(1,n)$ and large $n$ as seen in Figure~\ref{n20degen500} for $n=20$.

		\section{Conclusions}  	
	In this work we considered the algebraic structure of a special PT-symmetric semi-classical Rabi model. 		 
		It turns out that the condition for the existence of a finite-dimensional invariant subspace has the form of a ``resonance condition'' between the driving frequency and its constant term: $\nu_0/\omega=n+1$. 
		The time-dependent Schr\"odinger equation  then has a quasi-exact solution corresponding to a so-called ``exceptional point'' at the edge of the PT-symmetric region in parameter space \cite{Xie PRA 2018}. 
		The invariant subspace belongs to the spin-$(n/2)$ representation of $sl(2)$ \cite{Turbiner 88}, since the time-dependent Hamiltonian can be written in terms of $sl(2)$-generators acting in this representation.
		This allows to give determinantal expressions for the additional relation between the driving strength $\nu_1$ and the dissipative coupling $\gamma$ in closed form. 
		The trivial representation $n=0$ is a special case, in which the gain/loss term exactly balances the driving and leads to the simplest type of exceptional point. The quasi-exact solution is usually unique for a given parameter set $\{\nu_0/\omega,\nu_1/\omega,\gamma/\omega\}$, with the exception of the case with time-independent driving, $\nu_1=0$. Interestingly, some of the two possible Floquet solutions for fixed gain/loss $\gamma$ continue to exist (at least approximately) for $\nu_1\neq 0$ if the quotient $\nu_0/\omega$ is a large integer. Having two initial states with exact time development at resonance in the driven system may be of some value for applications of our semi-classical system in quantum information technology.   
		
		We observe that \eqref{Rab} falls into the class of double confluent Heun equations \cite{Ronveaux}. It has two irregular singular points of $s$-rank two at $z=0$ and $z=\infty$, which are created by a confluence of two pairs of regular singular points (the same class is sometimes called ``biconfluent'' \cite{slavyanov,Ishkhanyan}).  
		If the exponents of second kind vanish at both  $z=0$ and $z=\infty$, and thus allow a Frobenius expansion of the solution around $z=0$, it can be expressed as a polynomial in $z$ if certain constraints on the parameters are satisfied \cite{slavyanov,Ishkhanyan}.  
These restrictions are reduced in our case to the condition \eqref{npho} and the vanishing of the determinant of the matrix \eqref{matrix}. 

It seems that other quasi-exactly solvable models could also be a class of PT-symmetric models and provide a better and deeper understanding of the structure of the solutions. We are working to carry out a fairly complete study of this interesting exploration.

		\section*{Aknowledgments}
		The work of M.B. was supported by the Czech Science Foundation within the project 22-18739S. 
		D.B. was funded by the Deutsche Forschungsgemeinschaft (DFG, German Research Foundation), grant 439943572.
		The research of L.M.N. and S.Z. was supported by the Q-CAYLE project, funded by the European Union-Next Generation UE/MICIU/Plan de Recuperacion, Transformacion y Resiliencia/Junta de Castilla y Leon (PRTRC17.11), and also by projects RED2022-134301-T and PID2023-148409NB-I00, both funded by MICIU/AEI/10.13039/501100011033; financial support of the Department of Education of the Junta de Castilla y Leon and FEDER Funds is also gratefully acknowledged (Reference: CLU-2023-1-05).
		We are grateful for fruitful discussions with A.V. Turbiner during the preparation of this manuscript.


\begin{thebibliography}{99}
			
			\bibitem {Bender1998}
			C. M. Bender and S. Boettcher, Real spectra in non-Hermitian Hamiltonians having $\mathcal{PT}$-symmetry, Phys. Rev. Lett. \textbf{80} (1998) 5243.
			
			
			
			\bibitem {Bender2024}
			C. M. Bender and D. W. Hook, $\mathcal{PT}$-symmetric quantum mechanics, Rev. Mod. Phys. \textbf{96} (2024) 045002.
			
			\bibitem{Ruter}
			C. E. Rüter et al., Observation of parity–time symmetry in optic, Nat. Phys. \textbf{6} (2010) 192. 
			
			\bibitem{Konotop}
			V. V. Konotop, J. Yang and D. A. Zezyulin, Nonlinear waves in PT-symmetric systems, Rev. Mod. Phys. \textbf{88} (2016) 035002.
			
			\bibitem {NatureChang}
			L. Chang et al., Parity–time symmetry and variable optical isolation in active–passive-coupled microresonators, Nat. Phot. \textbf{8} (2014) 524. 
			
			\bibitem {NaturePeng}
			B. Peng et al., Parity–time-symmetric whispering-gallery microcavities, Nat. Phys. \textbf{10} (2014) 394. 
			
			\bibitem {NatureAshida}
			Y. Ashida, S. Furukawa and M. Ueda, Nat. Comm. \textbf{8} (2017) 15791. 
			
			\bibitem {NatureAssawaworrarit}
			S. Assawaworrarit, X. Yu and Sh. Fan, Robust wireless power transfer using a nonlinear parity–time-symmetric circuit, Nat. \textbf{546} (2017) 387. 
			
			
			\bibitem {NatureXiao}
			L. Xiao et al., Observation of topological edge states in parity–time-symmetric quantum walks 2017, Nat. Phys. \textbf{13} (2017) 1117. 
			
			
			\bibitem {NatureWeimann}
			S. Weimann et al., Topologically protected bound states in photonic parity–time-symmetric crystals, Nat. Mat. \textbf{16} (2017) 433. 
			
			\bibitem {NatChen}
			P. -Y. Chen et al., Generalized parity–time symmetry condition for enhanced sensor telemetry, Nat. Elec. \textbf{1 }(2018) 297. 
			
			\bibitem {NatureFan}
			S. Assawaworrarit and Sh. Fan, Robust and efficient wireless power transfer using a switch-mode implementation of a nonlinear parity–time symmetric circuit, 
			Nat. Elec. \textbf{3} (2020) 273. 
			
			\bibitem {NatureCao}
			W. Cao et al., Fully integrated parity–time-symmetric electronics, Nat. Nanotech. \textbf{17} (2022) 262. 
			
			\bibitem {NatureLiu}
			W. Liu et al, Floquet parity-time symmetry in integrated photonics, Nat. Comm. \textbf{15} (2024) 946. 
			
			
			\bibitem {NatureFritzsche}
			A. Fritzsche et al., Parity–time-symmetric photonic topological insulator, Nat. Mat. \textbf{23} (2024) 377. 
			
			
			\bibitem {Rabi}
			I. I. Rabi, Space quantization in a gyrating magnetic field. Phys. Rev. \textbf{51} (1937) 652.
			

			\bibitem{Xie 2013}
			Q. Xie et al., The quantum Rabi model: solution and dynamics, J. Phys. A: Math. Theor. \textbf{50} (2013) 113001.
				
			\bibitem{Lee}
			T. E. Lee and Y. N. Joglekar, PT-symmetric Rabi model: Perturbation theory, Phys. Rev. A \textbf{92} (2015) 042103.  
			
			\bibitem{Gong}
			J. Gong and Q. H. Wang, Stabilizing non-Hermitian systems by periodic driving, Phys. Rev. A \textbf{91} (2015) 042135.
			
			\bibitem{Li}
			J. Li et al., Observation of parity-time symmetry breaking transitions in a dissipative Floquet system of ultracold atoms, Nat. Comm. \textbf{10} (2019) 855.
			
			\bibitem{Liu}
			Y. Liu , L. Mao and Y. Zhang, Floquet analysis of extended Rabi models based on high-frequency expansion, Phys. Rev. A \textbf{105} (2022) 053717.
			
			\bibitem{Lu}
			X. Lu et al., PT -symmetric quantum Rabi model, Phys. Rev. A \textbf{108} (2023) 053712. 
			
			
			\bibitem {rich periodic structure phase-space crystal}
			L. Guo, M. Marthaler and G. Schön, 
			Phase space crystals: a new way to create a quasienergy band structure,
			Phys. Rev. Lett. \textbf{111} (2013) 205303. 
			
			\bibitem {quantum error correction}
			S. Mundhada et al., 
			Encoding a qubit in an oscillator,
			Phys. Rev. Appl. \textbf{12} (2019) 054051. 
			
			\bibitem {higher order squeezing}
			C. W. S. Chang et al.,
			Observation of three-photon spontaneous parametric down-conversion in a superconducting parametric cavity,
			Phys. Rev. X \textbf{10} (2020) 011011.
			
						\bibitem{Xie PRA 2018} 
			Q. Xie, Sh. Rong and X. Liu, Exceptional points in a time-periodic parity-time-symmetric Rabi model, Phys. Rev. A \textbf{98} (2018) 052122.
			

			
			\bibitem{Turbiner 88}  
			A.~Turbiner: Quasi-exactly-solvable problems and $sl(2)$ algebra,  Commun. Math. Phys.  \textbf{118} (1988) 467.
			
			\bibitem{Turbiner}
			A.V. Turbiner: One-dimensional quasi-exactly solvable Schrödinger equations, Phys. Rep. \textbf{642} (2016) 1. 
			
			\bibitem{artemio}  A.~Gonzalez-Lopez, N.~Kamran, and P.J.~Olver: Quasi-exact solvability,  Contemp. Math.  \textbf{160} (1994) 113.
			
			\bibitem{Ronveaux}
			A. Ronveaux (Ed.), \textit{ Heun’s differential equations}, Oxford University Press, New York (1995).
			\bibitem{slavyanov} S. Y. Slavyanov and W. Lay, \textit{Special functions: A Unified Theory based on Singularities}, Oxford University Press, New York (2000).
			
			\bibitem {Ishkhanyan}
			A. M. Ishkhanyan, T. A. Shahverdyan and T. A. Ishkhanyan, Thirty five classes of solutions of the quantum time-dependent two-state problem in terms of the general Heun functions, Eur. Phys. J. D \textbf{69} (2015) 10. 
			
			
		\end{thebibliography}
	\end{document}